\documentclass[conference]{IEEEtran}

\usepackage{comment}
\usepackage{graphicx}
\usepackage{tikz}
\usepackage{tikz-3dplot}
\usetikzlibrary{arrows.meta, positioning}

\usepackage{booktabs}
\usepackage{multirow}
\usepackage{enumitem}
\usepackage{amsmath}
\usepackage[framemethod=TikZ]{mdframed}
\usepackage[most]{tcolorbox}
\usepackage[hidelinks,bookmarks=false]{hyperref}

\begin{document}

%\title{Sustainability Antipatterns: A Unit of Analysis for Sustainable Software Engineering}
\title{No One Left Behind: Cross-Level Analysis for Sustainable Software Engineering}

%Filling the Gaps in Sustainability software-engineering research
\begin{comment}
    
\author{
\IEEEauthorblockN{Anonymous Author(s)}
\IEEEauthorblockA{

}
}
\end{comment}

\author{
%Anonymous Authors
\IEEEauthorblockN{Masoum Salehi}
\IEEEauthorblockA{
Anhalt University of Applied Sciences\\
Köthen, Germany\\
masoum.salehi@hs-anhalt.de
}
\and
\IEEEauthorblockN{Sandro Schulze}
\IEEEauthorblockA{
Anhalt University of Applied Sciences\\
Köthen, Germany\\
sandro.schulze@hs-anhalt.de
}
\and
\IEEEauthorblockN{Jacob Krüger}
\IEEEauthorblockA{
Eindhoven University of Technology\\
Eindhoven, The Netherlands\\
j.kruger@tue.nl
}
}

\maketitle

% ---- IEEE/ARXIV PREPRINT FOOTER ----
\makeatletter
\def\ps@IEEEtitlepagestyle{
  \def\@oddhead{}
  \def\@evenhead{}
  \def\@oddfoot{
    \parbox{\textwidth}{\footnotesize\textsf{© 2026 IEEE. Personal use of this material is permitted. Permission from IEEE must be obtained for all other uses, in any current or future media, including reprinting/republishing this material for advertising or promotional purposes, creating new collective works, for resale or redistribution to servers or lists, or reuse of any copyrighted component of this work in other works.}}
  }
  \def\@evenfoot{}
}
\makeatother
% ----------------------------------------------

\begin{abstract} 
Software engineers expect a software system to be efficient, maintainable, economically viable, and socially responsible throughout its lifecycle. 
Unfortunately, decisions made at the organizational, process, or product levels of software development---even when they improve one of these goals---often have unintended long-term consequences at other levels.
In response, software-engineering research has explored different ways to achieve \textit{software sustainability}, including energy efficiency, resource optimization, and code maintainability. 
However, existing research offers limited explanations for how sustainability problems emerge and reinforce one another across the socio-technical levels of software engineering. 
Moreover, we argue that sustainability challenges are not isolated but rather systemic: 
They stem from interactions among organizational priorities, development practices, and technical conditions. 
In this vision paper, we introduce the concept of \emph{sustainability anti-patterns} to capture conditions that systematically produce unsustainable outcomes. 
We demonstrate how this concept can help expose cross-level misalignments between socio-technical levels and sustainability dimensions that remain difficult to recognize through artifact-centric or dimension-specific analyses. 
To guide future research, we outline concrete directions for identifying, formalizing, detecting, and mitigating sustainability anti-patterns. 
We envision sustainability anti-patterns as a foundation for an integrated and actionable research agenda on sustainable software engineering.

\end{abstract}


\section{Introduction}
\label{sec:inro}

Software is crucial for almost all aspects of modern society. 
As such, the long-term sustainability of software is essential. 
Software sustainability is understood as a systematic and multidimensional concept that goes beyond energy optimization:
It also captures the technical durability, social equity, and economic viability of the software lifecycle~\cite{becker2015}. 
Yet, as recent studies argue~\cite{garcia2026,konig2025,Venters2018}, sustainability research in software engineering is in its infancy.
Critically, a multidimensional view is largely missing in empirical software-engineering research~\cite{chitchyan2016,mcGuire2023}. 

So far, research has evolved largely in parallel by considering dimensions with their inherent goals in isolation. 
For example, energy-efficiency research~\cite{georgiou2019,moises2018} aims to reduce the energy consumption and carbon footprint of software systems; technical sustainability research~\cite{avgeriou2016,Venters2018} emphasizes software quality; and research on developer well-being~\cite{penzenstadler2020,tanveer2021} examines human and organizational factors for sustainable practices. 
Research on each of these dimensions provides valuable insights, but is often done within distinct research communities that have own assumptions, metrics, and methodologies. 
As a result, existing research provides limited support for understanding how sustainability problems emerge, interact, and reinforce each other over time.

We argue that this limitation stems from a deeper mismatch between how sustainability has been studied and how software systems are developed. 
Software development is a socio-technical activity, meaning that it consists of organizational (social structures), process (work practices), and product (artifacts) levels~\cite{baxter2011}. %has organizational structures, processes, and technical decisions  %consists of  
These levels are interconnected: choices made at one level influence outcomes at others~\cite{tanveer2021}. 
For instance, organizational decisions to prioritize system delivery may lead to reduced testing as a process-level shortcut. 
In turn, complexity may increase, energy consumption can rise, and developers may become frustrated with the codebase. %These are not isolated issues but rather interconnected problems that span across socio-technical levels: technical problems (e.g., complexity, energy inefficiency), organizational problems (e.g., misaligned incentives, short-term priorities), and human problems (e.g., developer burnout). 
The question we want to pose for future research is: How can we understand and solve problems that span across socio-technical levels and affect multiple sustainability dimensions?
This question gains critical urgency with the rise of AI-assisted and agentic development, where fast-paced automation outpaces human oversight and lets sustainability problems emerge and compound unnoticed.

As a step in this direction, we introduce sustainability anti-patterns as a new concept for analyzing sustainable software engineering. 
A sustainability anti-pattern is a recurring pattern of interconnected socio-technical problems that collectively result in unsustainable outcomes. 
By capturing interconnections across both, socio-technical levels and sustainability dimensions, we provide a lens through which systemic sustainability problems can be made visible and actionable. %socio-technical levels and sustainability dimensions..
%This involves looking at the entire socio-technical system rather than individual dimensions in isolation. %individual parts
In this vision paper, we define the concept of sustainability anti-patterns, distinguish it from related concepts like traditional anti-patterns, illustrate its use through an example, and outline a research agenda for cataloging, detecting, measuring the impact on sustainability, and mitigating sustainability anti-patterns in software development.

\begin{figure*}
    \centering
    \includegraphics[width=0.90\linewidth]{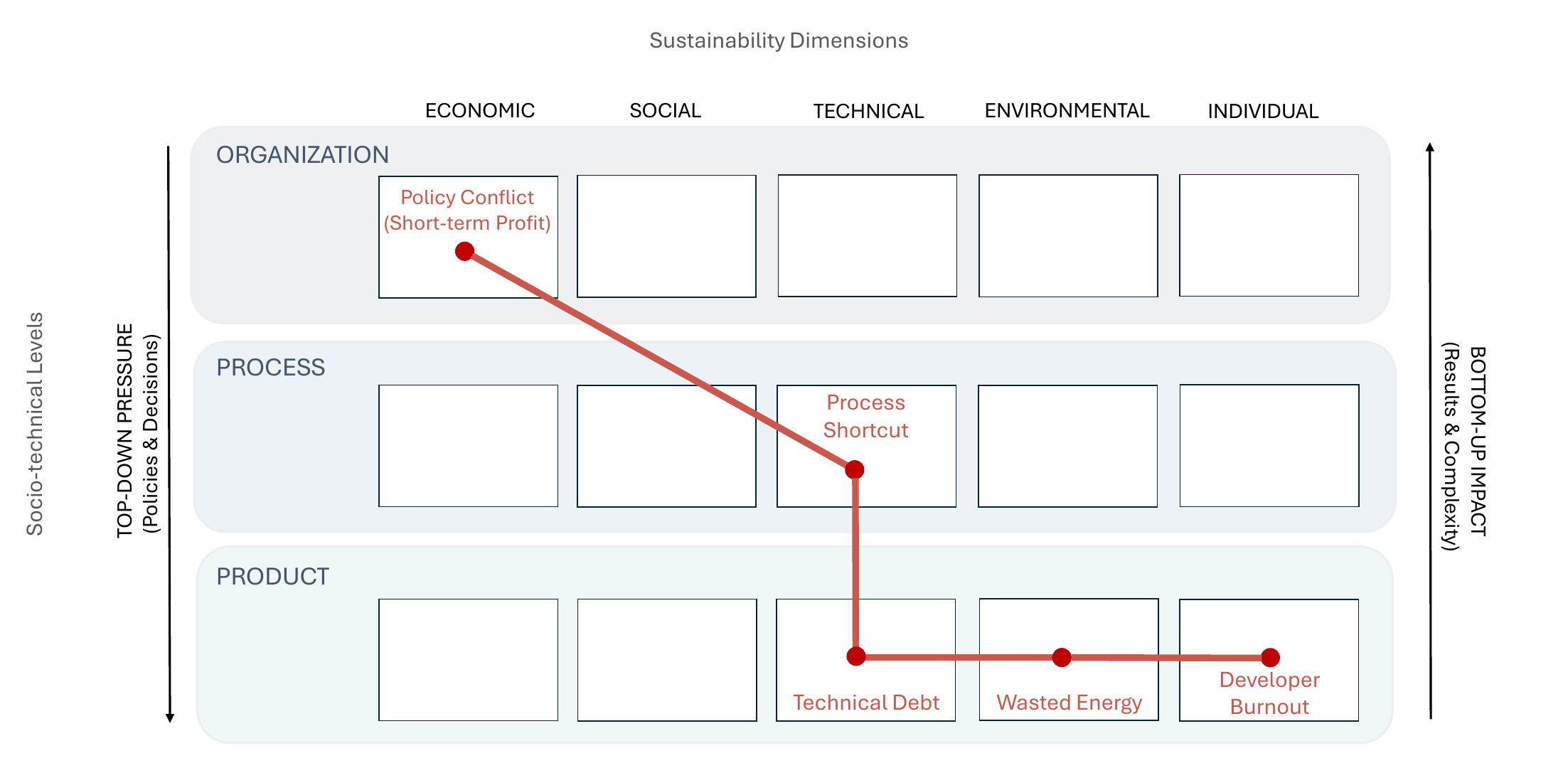}
    \caption{Exemplary overview of our socio-technical sustainability lens, illustrating how sustainability anti-patterns capture interactions across organizational, process, and product levels as well as across sustainability dimensions.}
    \label{fig:vision}
\end{figure*}

\section{The Need for Sustainability Anti-patterns}
\label{sec:gap}
Software-engineering research has made significant contributions to software sustainability since this topic gained popularity in 2010~\cite{noman2024}. 
However, current research is fragmented and does not fully capture the interconnected and recurring nature of sustainability problems. 
In this section, we discuss two major limitations that motivate our vision.

\subsection{Patterns and Anti-patterns}
Patterns are reusable, proven solutions to recurring problems within a given context. 
This concept was introduced in 1994 by Gamma et al.~\cite{gamma1995}, who documented recurring design solutions to common problems in object-oriented systems. 
In contrast, anti-patterns describe commonly repeated solutions that appear beneficial but ultimately produce negative consequences~\cite{brown1998}. Anti-patterns often look like good ideas initially (a solution) but lead to bad outcomes (negative consequences).

Several researchers cataloged patterns and anti-patterns across software development. 
For example, at the product level, code smells~\cite{fowler1999} indicate surface-level indicators of deeper design problems, such as duplicated code or long methods. 
At the process level, Brada and Picha~\cite{brada2019} have identified process anti-patterns, which undermine project management effectiveness. 
At the organizational level, community smells~\cite{tamburri2013} describe dysfunctional social structures---such as organizational silo or lone wolf---that hinder collaboration and knowledge sharing. 
However, these traditional patterns are typically studied in isolation and are rarely analyzed from a sustainability perspective. 
 
Even though recent research has introduced sustainability smells~\cite{kosbar2025}, energy smells~\cite{palomba2019}, and sustainability design patterns~\cite{lano2024}, these efforts focus on linking code and energy consumption. 
They still treat the problem as a technical optimization task. 
However, software systems' non-sustainability is systemic: multiple problems can co-occur and amplify one another. 
Subsequently, addressing a problem in isolation may fail if the underlying systemic configuration remains stable. The widespread reliance on technical debt~\cite{cunningham1992} illustrates this limitation clearly. 
Technical debt captures the accumulated cost of implementation shortcuts at the product level, representing an isolated artifact-centric symptom. 
However, it fails to account for the organizational conditions driving those shortcuts (cf. \autoref{fig:vision}), the process-level practices that allow debt to grow unchecked, or downstream sustainability consequences (e.g., developer burnout or wasted energy) that accumulate as a result. Addressing technical debt in isolation, without changing the cross-level socio-technical conditions that produced it, is precisely the kind of incomplete intervention that motivates our work.

\subsection{Systemic Nature of Sustainability}
Systems thinking is a comprehensive approach that focuses on how different components of a system interact and influence each other throughout time~\cite{richmond1993}. 
Software sustainability problems are, by their nature, complex system problems~\cite{penzenstadler2018}. %By sustainability problem, we mean any issue---whether at the organizational, process, or product level---that negatively affects one or more sustainability dimensions, often persisting or accumulating over time
%A sustainability problem does not reside in a single line of code, a single development practice, or a single organizational policy. 
%Rather, it emerges from the interplay of decisions across socio-technical levels and shows a direct or indirect impact on one or more sustainability dimensions. %levels%
In most cases of systemic interest, a sustainability problem cannot be fully explained by a single line of code, a single development practice, or a single organizational policy in isolation.
Rather, it emerges from---or is significantly amplified by---the interplay of decisions across socio-technical levels and shows direct or indirect impact on one or more sustainability dimensions.
Current research treats a chosen dimension and level as the primary area of analysis and rarely bridges the gap between organizational structures, development processes, and technical artifacts. 
Without such a holistic view, it remains unclear how local decisions propagate through the socio-technical levels of a project or how they reinforce each other across sustainability dimensions. %levels
The Karlskrona Manifesto~\cite{becker2015} as a foundation for sustainability design frames sustainability as such a systemic multi-dimensional concern. 
Other researchers have also acknowledged the systemic nature of sustainability~\cite{mcGuire2023,moreira2025}. 

Unfortunately,  empirical software-engineering research remains largely dominated by a reductionist paradigm, where the focus is on optimizing isolated system components, such as energy consumption or code maintainability. 
While this modularity is a cornerstone of traditional engineering, it is insufficient for sustainability, which is an emergent property of an entire socio-technical system. 
Some researchers have proposed systems thinking to improve sustainability in software engineering. 
Among others, finding leverage points as intervention targets for sustainability~\cite{penzenstadler2018}, requirements engineering as a systemic bridge~\cite{duboc2020}, or a capability maturity framework~\cite{sriraman2023} to assess an organization's software sustainability have been proposed. 
However, these efforts have remained largely theoretical and treat sustainability as static levels rather than dynamic behavior. 
Consequently, the software-engineering community still lacks a practical concept for identifying, cataloging, analyzing, and mitigating systemic sustainability problems in real-world projects.

\section{The vision: Sustainability Anti-patterns}
\label{sec:vison} 

The limited systemic view of sustainable software engineering we discussed in \autoref{sec:gap} motivates the need for a concept to systematically capture cross-level socio-technical relationships and sustainability dimensions. 
In Figure \ref{fig:vision}, we sketch an overview of our vision for moving towards such a systemic view based; including the example of technical debt. 
We introduce sustainability anti-patterns as a new concept for capturing recurring cross-level patterns, which collectively lead to unsustainable outcomes. 
Building on the notion of anti-patterns in software engineering~\cite{brown1998}, we define a sustainability anti-pattern as follows:

\newtcolorbox{definitionbox}[1][]{
    enhanced,
    colback=white,
    colframe=black,
    boxrule=0.5pt,
    arc=2mm,
    left=4pt,
    right=4pt,
    top=4pt,
    bottom=4pt,
    fonttitle=\color{black},
    title=#1,
    attach boxed title to top left={
        yshift=-2.2mm,
        xshift=4mm
    },
    boxed title style={
        colback=white,
        colframe=white,
        coltext=black,
        boxrule=0pt
    }
}
\begin{definitionbox}[Definition: Sustainability Anti-pattern]
A sustainability anti-pattern is a recurring socio-technical
pattern in which interacting organizational priorities,
development practices, and technical decisions intend to create
short-term local benefits while systematically generating
long-term unsustainable outcomes that span across
sustainability dimensions.
\end{definitionbox}

Our definition of sustainability anti-patterns extends traditional anti-pattern definitions~\cite{brown1998} in three important ways. 
First, it focuses on multi-level socio-technical patterns rather than isolated artifacts. 
Second, consequences are acknowledged to spread across sustainability dimensions.
Lastly, it highlights how seemingly rational short-term decisions can lead to long-term systemic failure. 
In our idea, such anti-patterns emerge from how choices, actions, and systems at different socio-technical levels connect to and influence one another. %levels
Specifically, this involves: (1) the organizational level (e.g., strategic priorities, incentive structures), (2) the process level (e.g., development practices, coordination mechanisms), and (3) the product level (e.g., code quality issues, technical debt). 

It is worth to note that our definition also captures how we intend to identify and address sustainability problems holistically. %problems
We envision to shift the focus towards the emergence of system-level effects from local decisions, rather than evaluating artifacts or practices in isolation. % decisions
This encourages researchers and practitioners to move beyond localized metrics towards examining interrelated patterns and their evolution over time. %patterns or %how local decisions interact and ..
However, to operationalize such patterns, it requires reasoning about complex dependencies rather than simple observations (cf. \autoref{sec:roadmap}). In the following, we illustrate the practical relevance of our vision through an example.

    \begin{figure}
    \centering
    \includegraphics[width=\linewidth]{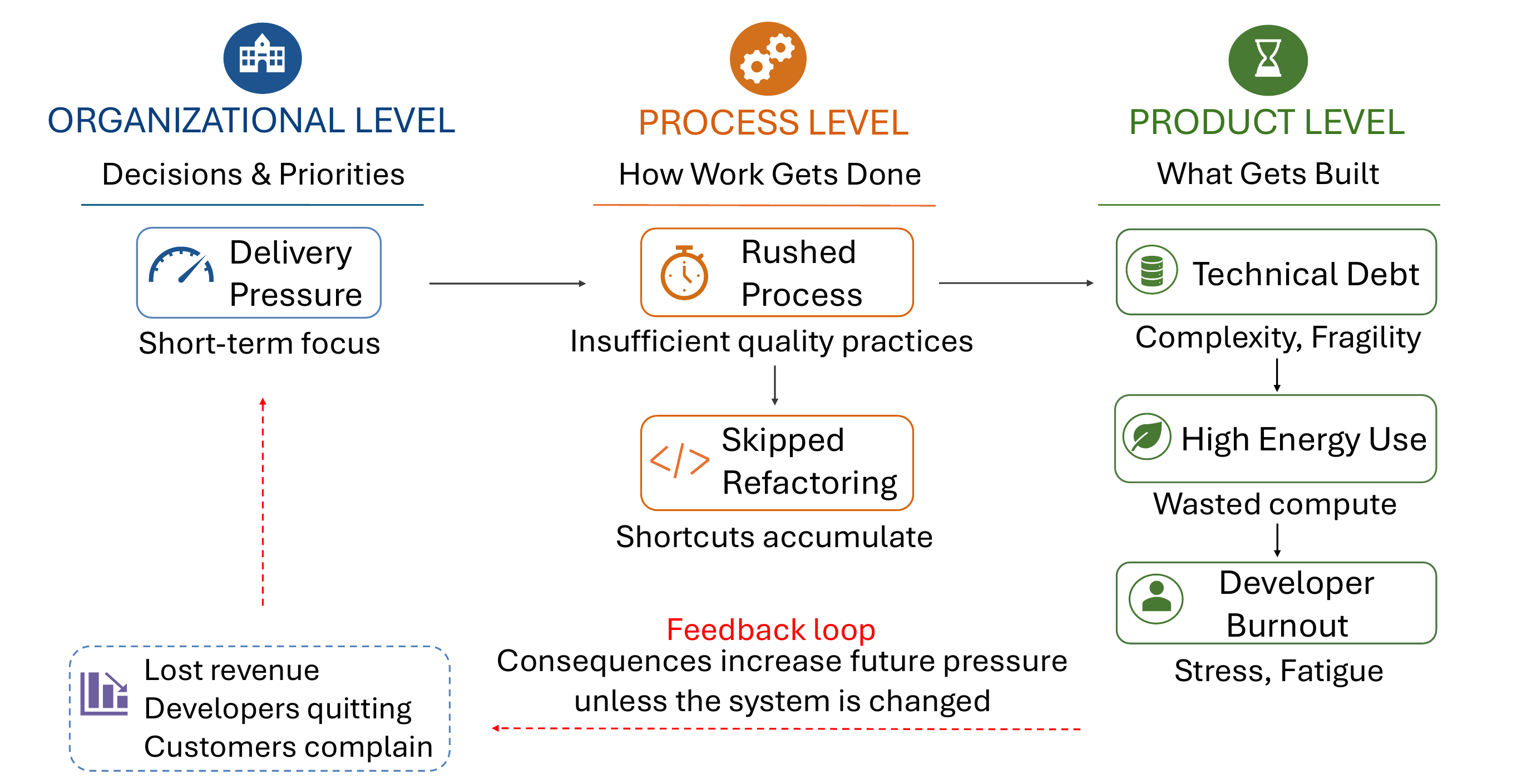}
    \caption{An example of how sustainability problems propagate across socio-technical levels and amplify.}
    \label{fig:example}
\end{figure}

% \subsection{Illustrative Example of Sustainability Anti-pattern}
In \autoref{fig:example}, we illustrate a high-level example of a sustainability anti-pattern. 
Note that this is not intended as an empirical finding, but as a conceptual instantiation that shows how local decisions across socio-technical levels may create unsustainable outcomes. %levels
Imagine a situation in software development where the organization's success is frequently measured in terms of short-term delivery metrics (e.g., feature throughput or release frequency). 
While this priority is often justified by market demands or competition, it can lead to decisions and practices at the process level that prioritize speed over rigor. 
For example, to meet tight deadlines, the development team may decide to reduce testing effort, skip refactoring, or minimize architectural discussions. 
This leads to technical debt at the product level \cite{avgeriou2016}---consciously or unconsciously, often with the expectation that it can be fixed later. 
Over time, the system becomes complex, making it more difficult to modify, evolve, and even understand. 
Subsequently, developers must invest more time and effort to implement changes, fix bugs, and ensure system stability. 
In some cases, inefficient implementations due to time constraints may also lead to higher resource consumption or poor performance \cite{palomba2019}. 
Eventually, developers may face burnout and turnover \cite{tanveer2021} while users experience declining product performance. 
Even worse, such effects do not remain local but propagate across levels and (sustainability) dimensions. 
In response, organizations may reinforce the original priorities, which results in a circle in which short-term optimizations continuously generate long-term costs.

% \noindent
% \textbf{Insights.} 
The example illustrates what a sustainability anti-pattern looks like and what it captures. 
Sustainability failures are not isolated technical issues, but rather cross-level misalignments between organizational priorities, development processes, and technical conditions. 
This structure produces unsustainable outcomes across multiple dimensions: technical (poor quality), environmental (energy waste), economic (increased maintenance cost), and human (developer burnout). 
While such individual outcomes have been researched in isolation, holistic anti-patterns, as in our example, have not been made explicit, yet.
We hypothesize that variants of such anti-patterns are common across software projects. 
However, in practice, they may vary in their specific causal paths, degrees of impact, and contextual triggers. 
We envision to identify, document, validate, and mitigate such patterns so that practitioners can recognize them and researchers can study them cumulatively.

\begin{comment}

    1. Should we give a name to the example?
    2. devil’s circle or self-reinforcing feedback loop which is more aligned with systems thinking? or even vicious cycle?
    3. Definition: a recurring socio-technical pattern? syndrome? configuration?
    4. The order in Section II, as it is or reverse?
    5. Title: I am not sure about:Cross‑Level Analysis. It feels more like a method than a concept
    6. Do we need to add page number?
    7. Name of authors: already solved.
    
\end{comment}

\section{Research roadmap}
\label{sec:roadmap}
To adopt sustainability anti-patterns as a lens for understanding software sustainability, we see several research opportunities and challenges. 
In the following, we outline key directions that together define an actionable agenda.

\noindent
\textbf{Understanding Sustainability Anti-Patterns.} 
The first step is to know how patterns of unsustainability actually look like in practice. 
In our vision, sustainability problems arise across socio-technical contexts and sustainability dimensions. 
This is grounded in systems thinking \cite{voulvoulis2022} and consistent with the socio-technical nature of software development \cite{baxter2011}, but the evidence is scattered across silos. 
We envision a systematic synthesis of existing literature to document currently separated sustainability concerns.
Understanding how these connect and reinforce each other remains an open problem. 
Systems-thinking methods, such as causal loop diagrams and the iceberg model, can enable researchers to hypothesize anti-patterns from findings that have already been published but never connected. 
Also, we call for direct field investigations. 
Empirical studies are essential, including practitioner interviews or software repository mining of process artifacts (e.g., issue trackers, CI logs) and organizational signals (e.g., team structures, workload distribution). 
Moreover, it is important to understand how the growing adoption of AI-assisted development and agentic systems creates new socio-technical dynamics that must be taken into account as well. 
We call for future work to investigate whether existing anti-pattern categories require extension or entirely new categories to account for these dynamics.

\noindent
\textbf{Formalization.} 
Since different researchers may describe the same anti-pattern differently, a shared, structured language for representing sustainability anti-patterns is important. 
Towards this direction, we need a formal template that specifies the essential elements of every anti-pattern. 
Such a template can be drawn from software design-patterns, but must capture socio-technical contexts and sustainability dimensions that design patterns are not intended to represent. 
We envision a structured template that captures, at least, an anti-pattern's name, socio-technical context, observable symptoms, impacted sustainability dimensions, and potential mitigation strategies. 
Equally critical is to define what a sustainability anti-pattern is not. 
%We suggest that a sustainability anti-pattern must satisfy two conditions: (1) recurrence across multiple projects and contexts and (2) impact on at least two sustainability dimensions. 
%When issues are strictly one-dimensional (e.g., a dip in technical performance), it is better to refer to technical debt and not require the anti-pattern framing. 
We suggest that a sustainability anti-pattern must satisfy two conditions: (1) recurrence across multiple projects and contexts and (2) an observable impact on more than one sustainability dimension. 
The first condition distinguishes sustainability anti-patterns from one-off project failures, while the second condition distinguishes them from dimension-specific problems, such as isolated technical debt. 
However, operationalizing both conditions remains a research challenge in itself, and the thresholds we propose are starting points subject to empirical refinements. 
When issues are strictly one-dimensional (e.g., a localized dip in technical performance without evidence of cross-dimensional propagation), existing concepts like technical debt or energy smells remain more appropriate framings.

\noindent
\textbf{Detection and Tooling.} 
To bridge the gap between research and practice, we need to make sustainability anti-patterns actionable for practitioners. 
This requires tool support.
However, a sustainability anti-pattern may manifest as measurable signals across different layers of socio-technical systems. 
Some indicators may reside in software repositories (e.g., code complexity trends, energy consumption profiles, test coverage decline, or coordination patterns). 
Other indicators exist only in practitioners' knowledge (e.g., organizational pressure, implicit team norms, or the real thoroughness of code). 
Yet, no single method currently examines them together. 
So, to capture a full anti-pattern, we propose combining repository mining with practitioner inquiries. 
For example, a structured self-assessment checklist could reference both repository metrics and human judgment, helping a team to check whether a known anti-pattern matches their project. 
Future tools must support continuous monitoring and analysis of sustainability anti-patterns. 
For example, dashboards and DevOps pipeline integrations could flag threshold violations before the underlying issues become systemic failures.

\noindent
\textbf{Intervention.} 
After detecting a sustainability anti-pattern, the ultimate goal is to fix it. 
In our idea, sustainability anti-patterns represent recurring patterns of interacting problems.
Thus, to address them, we require more than just local fixes (e.g., code refactoring). 
This means that we need effective interventions that may require changes at multiple levels at the same time. 
For example, to mitigate an anti-pattern that is caused by delivery pressure, we may need to align incentives, adjust development practices, and restructure parts of the system architecture. 
Future research must therefore develop and evaluate such mitigation strategies that match the socio-technical nature of each anti-pattern and understand potential trade-offs across sustainability dimensions. 
However, evaluating such strategies in real organizations is difficult, since some interventions are too risky to attempt, and sustainability outcomes may unfold only after months or years. 
Methods like longitudinal studies, quasi-experimental field studies, and simulation models may help researchers as a complementary way to explore intervention dynamics that are too slow or risky to study in real organizations. 

\noindent
\textbf{Sustainability Impact.} 
To know whether a remediation works, we must measure the impact anti-patterns have. 
However, this measurement is challenging because sustainability impacts accumulate across multiple sustainability dimensions on different timescales. 
For example, energy costs may increase over months, developer burnout may take years to manifest, and erosion of community trust leaves no digital trace. 
In addition, understanding how to separate the effect of one anti-pattern from other patterns is important. 
We propose longitudinal studies to compare similar projects over time. 
This will help researchers estimate what would have happened without the anti-pattern and to attribute identified impacts to specific patterns. 
Also, understanding who bears the costs is crucial, since sustainability harm is rarely distributed equally across stakeholders. 

\section{Conclusion} 

In this paper, we have introduced the concept of sustainability anti-patterns in software engineering. 
As we exemplified, sustainability anti-patterns are beyond localized problems; they are recurring socio-technical patterns that systematically produce long-term unsustainable outcomes across sustainability dimensions. 
Without a concept for analyzing such phenomena, current research is not able to explicitly capture cross-level interactions and will continue treating symptoms while underlying anti-patterns persist. 
We are convinced that, in addition to optimizing isolated problems (e.g., energy, maintainability), it is also important in software-engineering practice and research to understand how these problems interact and reinforce each other over time to enable systemic intervention. 
This need is amplified in the context of AI-assisted development and agentic systems, where the speed and automation of decision-making can accelerate the emergence of long-term unsustainable outcomes while also making them harder to trace.
As guidance, we outlined a research roadmap for identifying, formalizing, detecting, intervening on, and measuring the impact of sustainability anti-patterns. 
By tackling the research directions we have proposed, we aim to advance software engineering research and practice toward a more integrated and actionable understanding of sustainability.

% \bigskip
\noindent
\textbf{Acknowledgements}
This research is supported by the European Union under the project InterGrad-EGD (9110110301).

%\balance
\bibliographystyle{IEEEtranS}
\bibliography{ref2}

\end{document}